\documentclass[sigconf, acmsmall]{acmart}

\usepackage{rotating}
\usepackage{multirow}
\usepackage{graphicx}
\usepackage{xcolor}
\usepackage{array}

\newcommand{\change}[1]{\textcolor{black}{#1}}
\newcommand{\edit}[1]{\textcolor{black}{#1}}
\newcommand{\comments}[1]{\textcolor{black}{#1}}

\AtBeginDocument{%
  \providecommand\BibTeX{{%
    \normalfont B\kern-0.5em{\scshape i\kern-0.25em b}\kern-0.8em\TeX}}}

\setcopyright{acmcopyright}
\copyrightyear{2025}
\acmYear{2025}
\acmDOI{XXXXXXX.XXXXXXX}

\acmConference[CSCW '25]{Make sure to enter the correct
  conference title from your rights confirmation emai}{Nov 09-13, 2024}{San Jose, Costa Rica}
\acmPrice{15.00} % Need to change
\acmISBN{978-1-4503-XXXX-X/18/06} % Need to change

\begin{document}

%%
%% The "title" command has an optional parameter,
%% allowing the author to define a "short title" to be used in page headers.
\title[Barriers to expertise sharing among spreadsheet users]{"How do you even know that stuff?": Barriers to expertise sharing among spreadsheet users}

%%
%% The "author" command and its associated commands are used to define
%% the authors and their affiliations.
%% Of note is the shared affiliation of the first two authors, and the
%% "authornote" and "authornotemark" commands
%% used to denote shared contribution to the research.
\author{Qing (Nancy) Xia}
\affiliation{%
  \institution{University College London}
  \streetaddress{66-72 Gower Street}
  \city{London}
  \state{}
  \country{United Kingdom}
  \postcode{WC1E 6EA}
}
\email{nancy.xia.18@ucl.ac.uk}

\author{Advait Sarkar}
\affiliation{%
  \institution{Microsoft Research Cambridge; University College London; University of Cambridge}
  \country{United Kingdom}
}
\email{advait@emicrosoft.com}

\author{Duncan P. Brumby}
\affiliation{%
  \institution{University College London}
  \streetaddress{66-72 Gower Street}
  \city{London}
  \state{}
  \country{United Kingdom}
  \postcode{WC1E 6EA}
}
\email{d.brumby@ucl.ac.uk}

\author{Anna Cox}
\affiliation{%
  \institution{University College London}
  \streetaddress{66-72 Gower Street}
  \city{London}
  \state{}
  \country{United Kingdom}
  \postcode{WC1E 6EA}
}
\email{anna.cox@ucl.ac.uk}

% \author{Anonymous}
% \affiliation{%
%   \institution{}
%   \streetaddress{}
%   \city{}
%   \country{}
% \email{}
% }

% \author{Anonymous}
% \affiliation{%
%   \institution{}
%   \streetaddress{}
%   \city{}
%   \country{}
% \email{}
% }

% \author{Anonymous}
% \affiliation{%
%   \institution{}
%   \streetaddress{}
%   \city{}
%   \country{}
% \email{}
% }

%%
%% By default, the full list of authors will be used in the page
%% headers. Often, this list is too long, and will overlap
%% other information printed in the page headers. This command allows
%% the author to define a more concise list
%% of authors' names for this purpose.
\renewcommand{\shortauthors}{Anonymous}

%%
%% The abstract is a short summary of the work to be presented in the
%% article.
\begin{abstract}
Spreadsheet collaboration provides valuable opportunities for learning and expertise sharing between colleagues. \comments{Sharing expertise is essential for the retention of important technical skillsets within organisations}, but previous studies suggest that spreadsheet experts often fail to disseminate their knowledge to others. We suggest that social norms and beliefs \comments{surrounding} the value of spreadsheet use \comments{significantly influence user engagement in sharing behaviours}. To explore this, we conducted 31 semi-structured interviews with professional spreadsheet users from two separate samples. We found that spreadsheet \comments{providers face challenges in adapting} highly personalised strategies to often subjective standards and evaluating the appropriate social timing of sharing. In addition, \comments{conflicted self-evaluations of one's spreadsheet expertise, dismissive normative beliefs about the value of this knowledge, and concerns about the potential disruptions associated with collaboration can further deter sharing. We suggest these observations reflect the challenges of long-term learning in feature-rich software designed primarily with initial learnability in mind}. We therefore provide \comments{implications for design to navigate this tension}. Overall, our findings demonstrate how the complex interaction between technology design and social dynamics can shape collaborative learning behaviours in the context of feature-rich software.

\end{abstract}

%%
%% The code below is generated by the tool at http://dl.acm.org/ccs.cfm.
%% Please copy and paste the code instead of the example below.
%%


%%
%% Keywords. The author(s) should pick words that accurately describe
%% the work being presented. Separate the keywords with commas.

\received{20 February 2007}
\received[revised]{12 March 2009}
\received[accepted]{5 June 2009}

%%
%% This command processes the author and affiliation and title
%% information and builds the first part of the formatted document.
\maketitle

\section{Introduction}

For millions of users around the world, the spreadsheet is an essential tool for data handling and collaboration \cite{lawson2009, scaffidi2005, srinivasaragavan2021, serp2005}. Nearly 88\% from a sample of around 1600 individuals reported working on spreadsheets shared with at least one other user \cite{lawson2009}. Spreadsheet sharing activities, where viewing or editing rights are granted to another individual, are important for a variety of purposes. These include: ensuring data transparency, facilitating learning, enabling quality checks, and supporting teamwork \cite{srinivasaragavan2021, serp2005, sarkar2018}.

However, spreadsheet collaborations can also be vulnerable to errors, the consequences of which are often costly \cite{caulkins2007, eusprig2022, croll2007}. In certain circumstances, the likelihood of such errors can be exacerbated. The first is when collaborators have a mismatch in software-related expertise \cite{smith2017, srinivasaragavan2021}. Two of the five most frequently reported spreadsheet errors occur when users fail to detect mistakes, or fundamentally misinterpret the final output of inherited spreadsheets  \cite{caulkins2007}. The second is when organisations experience knowledge loss, such as when experienced users move on from the organisation without having shared their knowledge to others beforehand \cite{smith2017}. This is particularly problematic in the spreadsheet context, since often only the original spreadsheet creator possesses the unique technical, domain, and contextual knowledge needed to maintain and interpret the spreadsheet \cite{srinivasaragavan2021, kohlhase2015}. \comments{In such cases,} addressing the knowledge gap created is often laborious, \comments{and the remaining users are often forced to develop workarounds rather than developing the expertise needed to fill these gaps} \cite{smith2017}.

Developing users' technical skills is therefore crucial for reducing spreadsheet errors, particularly in professional contexts. While various learning approaches exist, we focus on knowledge/expertise sharing practices \cite{cockburn2014} among colleagues, such as over-the-shoulder learning \cite{twidale2005}. Peer interactions are widely recognised as one of the most important ways to learn about software usage \cite{nardi1991, kiani2020, lawson2009}.  Furthermore, this practice ensures retention of essential, work-related knowledge within an organisation, even when employees transition or transfer  \cite{smith2017}. 

\comments{However, designing interventions that effectively facilitate knowledge sharing is a complex problem}. Sharing interactions are not only multi-directional, but are also inherently social, and can involve multiple parties \cite{pipek2003, yang2019}. Typically, software-related expertise sharing is understood as a general form of knowledge sharing activity (e.g. answering questions) which happens to concern knowledge \textit{about} a software (e.g. \cite{joshi2020, kiani2020}). Our understanding of such sharing activities are well supplemented by previous research in the field of knowledge management. These works \comments{provide} key insights into how individual, social, and organisational factors (e.g. self-efficacy, social norms, rewards, and organisational culture) can affect expertise sharing behaviours \cite{ardichvili2003, riege2005}. 

In this paper, we highlight that certain types of feature-rich software—such as the spreadsheet—are embedded with their own specific social norms, expectations, and associated identities, even if they have not inherently been designed to evoke these associations \cite{sarkar2023}. These software-specific associations can have an important influence on user's engagement with productive, collaborative behaviours. This is well known in cases where technologies have been explicitly designed for social interactions (e.g. social media platforms \cite{calerovaldez2013}), where unique sets of etiquette and taboos can develop and shape the way people engage in the platform. We suggest similar norms and expectations also exist for professional-use software such as spreadsheets, and can serve as social barriers to expertise sharing practices. For example, expectations of ownership can affect \textit{when} and \textit{to what extent} it is acceptable to make changes in another user's spreadsheet \cite{church2023}. Similarly, beliefs about what distinguishes a spreadsheet ‘expert’ or a professional programmer from an `amateur’ can affect views on \textit{who} is allowed to critique another individual's spreadsheet \cite{sarkar2022}. 

\comments{We argue that paying greater attention to the socially constructed perceptions surrounding software expertise and acceptable usage practices can deepen our understanding of why people may, or may not, engage in expertise sharing. We apply this perspective to the spreadsheet context, with the goal of generating insights into the following questions:}

\textbf{RQ1:} What are the social barriers to expertise sharing among spreadsheet users? 

\textbf{RQ2:} How does the design and usage of spreadsheet systems contribute to the social dynamics of expertise sharing between users?

In our paper, we focus on spreadsheet sharing behaviours—\comments{the act of granting another user with physical or digital access to one's spreadsheets for viewing or editing purposes}. This behaviour was selected because research suggests that spreadsheet sharing can facilitate expertise sharing \comments{both explicitly and incidentally}. \comments{In other words,} while the primary goal of sharing may be collaboration, opening access to a spreadsheet often allows recipients to uncover new features and embedded practices, \comments{and therefore has important implications for learning \cite{sarkar2018}.}

Our work provides two main contributions to the CSCW community. \comments{First, we offer a detailed description of the key factors which deter spreadsheet expertise sharing. Through our interviews, we found that spreadsheet providers often struggle to adapt highly personalised use strategies to subjective standards of quality and to navigate the social timings for sharing. These challenges are compounded by concerns over potential collaboration errors, conflicted self-evaluations of expertise, and broader normative beliefs that devalue spreadsheet-related knowledge. Altogether, these findings highlight how feature-rich software amplifies the social challenges of sharing due to their inherent flexibility, technical complexity, and potential ambiguity in how contents and functions are perceived.}

The second contribution of our work is a set of design suggestions \comments{which aim to mediate the aforementioned challenges} and promote knowledge sharing in the spreadsheet context. We link our observations to previous discussions around designing feature-rich software for ease of learnability, and build on previous literature to highlight how different software design choices may produce different social impacts and user dynamics in the long term.

\section{Related works}
\subsection{Spreadsheets as `praxisware'}
Designing for software learnability has long been a topic of core interest in the field of Human-Computer Interaction \cite{grossman2009a}.  While the field typically focuses on initial adoption and accessibility, designing for long-term learning in feature-rich software presents unique challenges \cite{grossman2009a, sarkar2023}. We borrow the term `praxisware' from Sarkar \cite{sarkar2023} to discuss the types of software for which the issue of long-term learnability is particularly salient. Praxisware are distinguished from general-use software in three ways: 1) the richness and diversity of functionalities which they provide; 2) the longevity of their usage—typically spanning years or even the entire length of an individual's career; and 3) being deeply embedded within specific professional identities and communities, unlike general-use software \cite{sarkar2023}. In other words, they are tools which both shape and allow users to express their professional skills, values, and identity (e.g. Photoshop, AutoCAD) in a way that is not encapsulated \comments{in} other types of software (e.g. Microsoft Word, PowerPoint)\cite{nouwens2018}. This incentivises learning, as development of one's technical expertise is not only practical, but \comments{can} also have important personal, social, and professional implications \cite{lemmetty2020}. 

In this paper, we consider spreadsheets to be a type of praxisware. Firstly, spreadsheets are feature-rich, offering a wide range of functionalities that can be challenging to master without long-term investment and support from others \cite{hermans2016, sarkar2018, smith2017}. Secondly, spreadsheets are strongly affiliated with professions such as finance, economics, and banking \cite{lawson2009, croll2007}, so high levels of spreadsheet expertise are often closely associated with high levels of professional expertise \cite{nouwens2018}. This means the study of spreadsheet-related knowledge sharing will have implications for both software learning and professional knowledge sharing contexts. Finally, previous works suggest that a `spreadsheet user' is in itself a marked social identity \cite{brekhus1998} with specific social associations and traits distinct from related identities such as `programmers' \cite{sarkar2022, hermans2016}.  This suggests that social norms and professional implications related to spreadsheet learning and expertise could influence engagement with knowledge sharing practices. By studying spreadsheet software, we can provide valuable insights into how attitudes, norms, and social identities around spreadsheet use could influence collaborative behaviours.

\subsection{Supporting software learning through expertise sharing}
\label{subsec: software_learning_lit}
Colleagues play a crucial role in supporting \comments{expertise development} in software use \cite{lemmetty2020, twidale2005, kiani2020}. Studies show that learning from colleagues through conversations and demonstrations \comments{happen more often and is more} preferred over alternative \comments{resources such as} search engines, training courses, help pages, \comments{or} videos \cite{kiani2020, lawson2009, lemmetty2020}.

Given the prevalence of interpersonal learning, researchers have proposed \comments{various} technical innovations to help \comments{users} better leverage colleagues' expertise \comments{for learning purposes. A key tension which researchers must consider is the balance between the depth of expertise shared, and the workload of the helper \cite{giannisakis2022, larsen-ledet2022}. Some tools prioritise giving user access to experienced individuals when needed. For example,} expert recommendation systems facilitate searches for knowledgeable individuals in an organisation by generating profiles about other users \cite{ackerman1990, riahi2012, paul2016, schoegje2024, husain2019}. Tools \comments{such as} `MarmalAid' \cite{chilana2018} \comments{and} `MicroMentor' \cite{joshi2020} \comments{support the communication process} by providing easy access to individual experts for \comments{synchronous} in-context help. \comments{Other approaches focus on reducing} the effort associated with capturing and documenting knowledge for sharing (e.g. \cite{chen2024, grossman2010, lafreniere2011}), such as by using crowd-sourcing approaches (e.g. \cite{bunt2014, lafreniere2013, dubois2017, matejka2009, lafreniere2011, matejka2013}). \comments{The goal is to make} knowledge sharing interactions more easily accessible and less effortful for sharers, \comments{therefore increasing possible learning opportunities for recipients}. 

While \comments{the above} research \comments{helps to identify possibilities for} how expertise sharing activities can be supported, previous studies often implicitly assume that knowledgeable individuals are always willing to share or reveal their expertise. However, \comments{there are also} many undesirable social repercussions associated with publicising one's expertise in the workplace \cite{larsen-ledet2022, pipek2003, giannisakis2022}. For example, individuals who are deemed `more' experienced than others may be disproportionately allocated responsibilities or expected to meet unrealistic demands \cite{diaz2020, wu2022a}. Giving other users visibility over one's software usage metrics or a means to quantify their own expertise can lead to concerns of surveillance, fears of criticism, and social comparison with others \cite{giannisakis2022, pipek2003}. Additionally, as Pipek et al. \cite{pipek2012} point out, in professional contexts, the role of the learner and helper are not mutually exclusive or binary, and one person often occupies the role of both interchangeably depending on the context. However, if an individual is unable to transition from a learner to helper, or recognise how their expertise may support others, then they are unlikely to engage in knowledge sharing activities—even if they are sufficiently equipped.

The effort of sharing is therefore not the only barrier to knowledge sharing. Identifying the social and individual consequences of sharing, from the knowledge sharer's perspective, is also essential. Our goal is to inform the design of spreadsheet technologies based on that perspective.

\subsection{\change{Knowledge management and contributors to knowledge sharing}}
\label{subsec:KM}

\comments{`Knowledge management' is the field of study on how} knowledge, as a type of intellectual asset, should be retained and cultivated within organisations \cite{jennex2006, baskerville2006}. An integral process in the management of these intellectual assets is the process of knowledge/expertise sharing \cite{baskerville2006, wang2010}. There are two common ways in which the process of sharing is defined \cite{ackerman1990, martensson2000}: 1) sharing `knowledge' through artefacts, by explicitly formalising knowledge through documentation or storing in an accessible database; 2) sharing `expertise' via interpersonal communication, which supports the transfer of information that is inherently challenging to document (e.g. personal and practical experience, intuition, know-how \cite{nonaka1995a}). \comments{As both documented knowledge and tacit expertise are relevant for feature-rich software operation,} we use both terms interchangeably in this paper.

Existing research from knowledge management has sought to provide a comprehensive understanding of the \comments{motivating factors} influencing knowledge sharing intention and behaviour \cite{wang2010, fauzi2024}. Theoretical frameworks of behaviour and decision-making, such as the theory of social exchange and social capital \cite{zhao2023}, social cognitive theory \cite{bandura2001}, or the theory of planned behaviour (TPB) \cite{ajzen1985, ajzen1991}, are frequently applied in survey-based studies to support researchers in identifying the relationships between relevant determinants of sharing behaviours (for reviews, see \cite{wang2010, thomas2021a, fauzi2024}). 

Such studies establish an important foundation for understanding the processes \comments{underlying} knowledge sharing in the workplace. For example, cultivating cultures which establish pro-sharing norms \cite{kankanhalli2005, bockgee-woo2005}, fostering a collaborative rather than a competitive environment \cite{willem2006}, and reducing the emphasis on rank and hierarchy \cite{mintzberg1980} can facilitate information flow between individuals and groups. In contrast, where knowledge sharing may be viewed as a loss of an individual's own unique expertise or asset \cite{szulanski1996}, or where there may be challenges associated with impression management (e.g., if knowledge providers are seen to be self-serving or disagreeable through the act of sharing) \cite{bolino1999}, knowledge sharing intention is often lower. Interpersonal relationships, such as the degree of trust between individuals \cite{kankanhalli2005}, as well as an individual's own characteristics, such as self-efficacy \cite{hsu2007, bock2002}, \comments{can also affect a person's willingness to engage in sharing activities.}

Another key focus in knowledge management is the use of technologies to facilitate the \textit{communication} of knowledge. Previous works have applied frameworks such as \comments{TPB} to explore how online/virtual platforms and social media facilitate knowledge sharing activities (for reviews, see \cite{fauzi2019, fauzi2024}). These studies demonstrate that the quality of a system (e.g., its ease of use, its accessibility) can positively affect sharing behaviours by increasing users' perceived behavioural control (i.e. their belief in their ability to execute a desired behaviour \cite{ajzen1985, ajzen1991}) and their perception of resource availability (e.g. feeling like they have more time available) \cite{hsu2008, lin2020a}. However, while effective system designs can be useful in producing positive sharing experiences, embedding the social nuances and motivators effectively into technological systems can be challenging. For example, implementing virtual social rewards (e.g. social points for answering questions on a forum) can, in some cases, undermine the positive feelings of altruism, enjoyment in helping others, and desire to build connections, which otherwise predict knowledge sharing on these platforms \cite{zhao2016, wang2022a}.

It is therefore essential that technology designers and researchers clearly understand the social influences underlying \comments{interactions among users} if we are to design systems that effectively promote expertise sharing. \comments{To contribute to that understanding, our study aims to identify how associated norms in spreadsheet-specific practices influence expertise-sharing intentions.}

\subsection{Opportunities and challenges for expertise sharing in spreadsheet collaborations}
\label{subsec: spreadsheet_collab}
Several prior works have explored knowledge sharing \comments{activities around spreadsheets. Research suggests that, alongside numerical data, a spreadsheet records the traces of how an analysis is conducted, the history of decision-making and intentions \cite{chalhoub2022, srinivasaragavan2021, kohlhase2015}, as well as contextual labelling and interpretation of information \cite{kankuzi2016} \change{(e.g. `Department Budget 2024')}}. Furthermore, spreadsheet sharing triggers interpersonal communication, encouraging users to externalise tacit knowledge via annotations, comments, and edits \comments{within or around a spreadsheet file} (e.g. via in-person help-seeking, emails, or meetings) \cite{srinivasaragavan2021, chalhoub2022, kohlhase2015, sarkar2018}. 

\comments{Research focusing on spreadsheet has explored the challenges and opportunities for spreadsheet-related knowledge sharing. One key finding is that internal organisational learning from peers is particularly valuable as users are not always at liberty to share confidential information in spreadsheets outside the organisation \cite{smith2017}.} Conversations with other users and help-seeking interactions are also commonly reported methods for self-learning among spreadsheet users \cite{sarkar2018, lawson2009}. Indeed, the absence of these interactions and a failure to transfer knowledge to relevant parties can result in severe knowledge gaps within the organisation \change{which can negatively impact performance} \cite{smith2017}. It is therefore important to develop a greater understanding of the social processes which may deter users from effectively utilising these opportunities for learning. 

Prior research suggests that the technical complexity required to find and interpret information in spreadsheets can be a barrier to such forms of knowledge transfer, unless users directly communicate and clarify with each other \cite{srinivasaragavan2021}. \comments{Common technical issues exacerbating the problem include}: the issue of visibility and presentation—users must interact directly with cells in order to reveal the underlying formulae or to structure the data in a way which is meaningful \cite{srinivasaragavan2021, chalhoub2022, bartram2022}; issues of logic interpretation and implementation—particularly when confronted with logically complex or visually dense formulae \cite{hermans2011, hermans2012a, smith2017, panko2016, yang2020}; or issues with the language of the interface—\comments{which are often based on} spreadsheet-terminology (e.g. `cell A1') \comments{rather than representations clear to users or their specific domains} (e.g. `profit,' `budget') \cite{kankuzi2016, cunha2020, hermans2011, hendry1993}.

Findings of spreadsheet research \comments{suggest} that documenting and extracting tacit knowledge may be too effortful. Users may therefore \comments{perceive the required resources (e.g. time, attention) as too high for knowledge sharing to be worthwhile.} However, to our knowledge, there has been little to no explicit focus in the current research on the social norms, attitudes, and beliefs around spreadsheet-related knowledge sharing and why users may or may not choose to engage in these practices. \comments{Our study aims} to address this gap. 

\section{Method}

To address the research questions previously stated, we analysed the interview data collected in two separate studies, originally designed for another research exploration. We decided to use this data to investigate the current questions for two reasons. First, interviews in both studies included sections dedicated to topics such as spreadsheet collaboration, help-seeking and help-giving behaviours, practices used to support spreadsheet comprehension when sharing, and discussions of industry standards and social norms of spreadsheet use. Second, interview sets from each study offered complementary perspectives advantageous for triangulating the findings. Study 1 provided a general overview of the social, organisational, and industry-based contexts surrounding spreadsheet use and learning. Study 2 provided insights into how users applied, discovered, explored, and communicated about different aspects of the spreadsheet (with a specific focus on formulas) when working with others. 

In the following sections, we broadly outline the original aims, recruitment strategies, and materials of each study. We place our emphasis on materials directly relevant to the questions in this paper. Ethical procedures of these studies are included in Appendix D.

\subsection{Study 1}
\subsubsection{Original aims}
Study 1 was conducted between February and March 2023. The original goal was to explore how users' interactions, practices, learning, and mental models of the spreadsheet differed according to their professions. It was therefore important that participants were recruited from a variety of professional backgrounds.

To ensure sample diversity, \comments{the chosen} professional domains from which we invited participants had to be sufficiently distinct in terms of their spreadsheet practices and resources for spreadsheet learning. For that, we evaluated the domains based on five key criteria that we identified:

\begin{itemize}
    \item \textit{Organisational structure} (e.g. whether the nature of an individual's role is team-based or predominantly driven by the individual);
    \item \textit{Prevalence of spreadsheet use} (i.e. the estimated proportion of time dedicated to spreadsheet-related work within working hours);
    \item \textit{Required spreadsheet expertise for domain} (the expected difficulty, commitment, and level of knowledge and mastery of spreadsheets necessary to carry out daily tasks in the field);
    \item \textit{Relevance of field-specific knowledge to spreadsheet use} (e.g. an individual possessing knowledge of programming languages may be more proficient at using spreadsheet formulas compared to an administrator);
    \item \textit{Prevalence of organisational support for spreadsheet use} (e.g. the availability of training sessions, workshops, guidelines, online or academic courses, informal support or 'buddy' systems etc
\end{itemize}

We ultimately identified four professional domains (technical, financial/data science, scientific, and administrative) to recruit participants. A detailed evaluation of these domains in terms of the five criteria is included in Appendix A.

\subsubsection{Participants}
We identified participants in these four selected domains through personal connection, professional social networks (e.g. LinkedIn), and a pre-study survey distributed on the online participant recruitment platform Prolific. Though the use of online platforms can pose risks to the validity of participant responses, Prolific participants largely produce higher quality data and engagement compared to rival platforms \cite{douglas2023, peer2022}.

All participants filled out a 10-minute pre-study recruitment survey, which allowed us to verify and ensure diversity in their professional background, spreadsheet usage, and demographics, before they were invited to the interview. Overall, 62 participants filled out the survey, 17 of whom then proceeded to the interview. Four participants were recruited for three of the four professional domains. For the administrative domain, five participants were recruited. The demographics of these 17 participants are shown in Table \ref{tab:S1_p_demog}.

\begin{table*}
  \centering

  \caption{Demographics of Study 1 interview participants}
  \label{tab:S1_p_demog}
  
  \resizebox{14cm}{!}{
  \begin{tabular}{
    >{\centering\arraybackslash}p{1.2cm}   % ID
    >{\centering\arraybackslash}m{2.8cm}   % Age (Gender)
    >{\raggedright\arraybackslash}m{4.2cm} % Occupation
    >{\centering\arraybackslash}m{3cm}     % Seniority
    >{\raggedright\arraybackslash}m{5.8cm} % Spreadsheet expertise
  }

    \multicolumn{5}{c}{\textbf{Technical domain}} \\
    \textbf{ID} & \textbf{Age (Gender)} & \textbf{Occupation} & \textbf{Seniority} & \textbf{Self-reported spreadsheet expertise} \\
    \hline
    
    S1P5 & 18–29 (M) & IT consultant & Non-manager & Extensive experience, some expertise \\
    S1P32 & 40–49 (M) & Data migration specialist & Supervisor/Manager & Some experience, still a beginner \\
    S1P39 & 40–49 (M) & Self-employed IT tester & Non-manager & Extensive experience, some expertise \\
    S1P15 & 60+ (M) & Independent app developer & Executive & Extensive experience, some expertise \\[1ex]

    \multicolumn{5}{c}{\textbf{Financial/Data science domain}} \\
    \textbf{ID} & \textbf{Age (Gender)} & \textbf{Occupation} & \textbf{Seniority} & \textbf{Self-reported spreadsheet expertise} \\
    \hline
    
    S1P1 & 18–29 (M) & Health economist & Non-manager & Very experienced, high expertise \\
    S1P3 & 18–29 (M) & Consultant & Non-manager & Extensive experience, some expertise \\
    S1P4 & 18–29 (F) & Data scientist/Researcher & Non-manager & Extensive experience, some expertise \\
    S1P21 & 40–49 (F) & Head of Treasury & Supervisor/Manager & Very experienced, high expertise \\[1ex]
    
    \multicolumn{5}{c}{\textbf{Scientific domain}} \\
    \textbf{ID} & \textbf{Age (Gender)} & \textbf{Occupation} & \textbf{Seniority} & \textbf{Self-reported spreadsheet expertise} \\
    \hline
    
    S1P26 & 18–29 (M) & Biomedical researcher & Non-manager & Extensive experience, some expertise \\
    S1P29 & 40–49 (F) & Biochemistry researcher & Supervisor/Manager & Extensive experience, some expertise \\
    S1P34 & 18–29 (M) & Neuroscientist & Non-manager & Some experience, still a beginner \\
    S1P45 & 18–29 (M) & Chemistry research assistant & Non-manager & Some experience, still a beginner \\[1ex]

    \multicolumn{5}{c}{\textbf{Administrative domain}} \\
    \textbf{ID} & \textbf{Age (Gender)} & \textbf{Occupation} & \textbf{Seniority} & \textbf{Self-reported spreadsheet expertise} \\
    \hline
    
    S1P2 & 18–29 (F) & Sales and distribution & Non-manager & Some experience, still a beginner \\
    S1P22 & 40–49 (F) & Auditor & Supervisor/Manager & Some experience, still a beginner \\
    S1P23 & 18–29 (M) & Marketing consultant & Executive & Extensive experience, some expertise \\
    S1P24 & 30–39 (F) & Human resources & Supervisor/Manager & Extensive experience, some expertise \\
    S1P25 & 30–39 (F) & Facilities manager & Supervisor/Manager & Extensive experience, some expertise \\
    
  \end{tabular}
  }
\end{table*}

\subsubsection{Materials}
The study involved a pre-study survey for recruitment and an interview guide. The survey was a shortened form of the Spreadsheet Engineering Research Project (SERP) survey, a well-established tool used to summarise spreadsheet practices in organisations \cite{lawson2009, baker2006}. We condensed the SERP survey from its original 67 questions to 20, as our main goal was to capture broad overviews of participants' spreadsheet practice which we then used to generate in-depth follow-up questions during the interviews. We therefore did not want participants to dedicate excessive amounts of time to the survey. We selected one or two relevant questions from each section of the SERP survey, focusing on participants' demographics, professional backgrounds, self-reported spreadsheet expertise, purpose of spreadsheet use, time spent on spreadsheet-related work, size and complexity of spreadsheets, number of spreadsheet collaborators and sharing practices, organisational policies, and frequently used spreadsheet learning resources. Topics which were deemed less relevant for recruitment purposes and more suited to interview explorations (e.g. spreadsheet risk management) were excluded. We also added 6 questions exploring whether participants are self-taught as well as their familiarity with a range of spreadsheet features, experience with programming languages, and willingness to participate in the interviews.

The interview guide comprised two sections. The first section was a general discussion addressing three main areas. The first area was about clarifying survey responses and gathering information on participants' work responsibilities and spreadsheet usage. The second examined spreadsheet-related practices, including team collaboration and spreadsheet-sharing habits. The final area focused on spreadsheet learning, user expertise, and help-seeking practices.

The second section of the interview guide prompted users to share and briefly describe an example spreadsheet they used at work. We first asked questions about the specific application of features for this particular spreadsheet (e.g. the use of formulas, data layout, visualisation strategies). We then asked participants to explain the decisions behind the creation of the spreadsheet. Participants were then encouraged to reflect on the challenges they faced in spreadsheet use and to provide suggestions for future tools to address these challenges (see Appendix B for the Study 1 survey and interview guide).

\subsubsection{Procedure}
A survey link was shared with participants on our social networks and Prolific. Participants who filled out the survey were invited on a rolling basis, and they booked their own timeslot. The interviews lasted around 1 hour, and were all conducted, recorded, and transcribed online by the first author using Microsoft Teams, an online video meeting platform, and its built-in transcription services. The order of questions were kept flexible to ensure a natural flow around the participants’ responses. Participants were first asked a set of follow-up questions based on the survey. They were then brought into the interview sections for an in-depth discussion. Finally, they were invited to screen-share and discuss their example spreadsheets.

\subsection{Study 2}
\subsubsection{Original aims}
Study 2 was originally conducted between August and October 2018. It focused on understanding how users interact with spreadsheet formulas, which are an important aspect of spreadsheet use, and how they developed expertise in using them. Because the study prioritised interaction, learning experience, and the factors affecting users’ engagement with spreadsheet formulas, recruitment was based on diversity in self-reported spreadsheet expertise and programming expertise. 

\subsubsection{Participants}
A total of 14 participants were recruited from the research organisations and partners we are affiliated with as well as from a pool of participants engaged in our previous research projects. The demographics of these participants are outlined in Table \ref{tab:S2_p_demog}.

\begin{table}
    \centering
    \captionof{table}{Demographics of Study 2 participants}
    \label{tab:S2_p_demog}

    \resizebox{14cm}{!}{
    \begin{tabular}{
      >{\raggedright\arraybackslash}p{0.75cm}  % ID
      >{\centering\arraybackslash}m{6cm}  % Domain
      >{\centering\arraybackslash}m{8cm} % Expertise
    }
    \hline
    \textbf{ID} & \textbf{Domain} & \textbf{Self-reported spreadsheet expertise} \\
    \hline
    S2P1 & Spectrum systems engineer & A lot of experience, and I use some advanced features \\
    S2P2 & Data scientist & A lot of experience, and I use some advanced features \\
    S2P3 & Business operations specialist & A lot of experience, and I use some advanced features \\
    S2P4 & Teaching administrator & A lot of experience, and I use some advanced features \\
    S2P5 & Commercial real-estate finance analyst & A lot of experience, and I use some advanced features \\
    S2P6 & Machine learning \& data science student & A lot of experience, and I use some advanced features \\
    S2P7 & Financial analyst & A lot of experience with many advanced features \\
    S2P8 & HCI PhD student & A lot of experience, but my use is basic \\
    S2P9 & Lawyer/MBA student & Some experience, but I'm still a beginner \\
    S2P10 & Data scientist & A lot of experience, but my use is basic \\
    S2P11 & Product manager & A lot of experience, and I use some advanced features \\
    S2P12 & Centre co-ordinator & A lot of experience, but my use is basic \\
    S2P13 & Civil servant & A lot of experience, but my use is basic \\
    S2P14 & Statistics student/ex-MBA analyst & A lot of experience, and I use some advanced features \\
    \hline
    \end{tabular}
    }

\end{table}

% \begin{table}
%     \centering
%     \captionof{table}{Demographics of Study 2 participants}
%     \label{tab:S2_p_demog}

%     \resizebox{14cm}{!}{
    
%     \begin{tabular}{p{0.5cm}c{3cm}p{4cm}}
%     \hline
%     \textbf{ID} & \textbf{Domain} & \textbf{Self-reported spreadsheet expertise} \\
%     \hline
%     S2P1 & Spectrum systems engineer & A lot of experience, and I use some advanced features\\
%     S2P2 & Data scientist & A lot of experience, and I use some advanced features\\
%     S2P3 & Business Operations Specialist & A lot of experience, and I use some advanced features\\
%     S2P4 & Teaching administrator & A lot of experience, and I use some advanced features\\
%     S2P5 & Commercial real-estate finance analyst & A lot of experience, and I use some advanced features\\
%     S2P6 & Student—Machine Learning and Data Science & A lot of experience, and I use some advanced features\\
%     S2P7 & Financial analyst & A lot of experience with many advanced features\\
%     S2P8 & PhD student in HCI & A lot of experience, but my use is basic\\
%     S2P9 & Lawyer/MBA student & Some experience, but I'm still a beginner\\
%     S2P10 & Data scientist & A lot of experience, but my use is basic\\
%     S2P11 & Product manager & A lot of experience, and I use some advanced features\\
%     S2P12 & Centre co-ordinator & A lot of experience, but my use is basic\\
%     S2P13 & Civil servant & A lot of experience, but my use is basic\\
%     S2P14 & Statistics student/ex-MBA analyst & A lot of experience, and I use some advanced features\\
%      \end{tabular}

%     }

% \end{table}

\subsubsection{Materials}
Similar to Study 1, Study 2 also involved a short pre-study survey for recruitment, and an interview guide. The survey captured the key demographics of the participants, including occupation, self-reported spreadsheet experience, spreadsheet use purpose, as well as expertise in both spreadsheet formulas and programming languages.

The design of the interview guide focused specifically on the barriers encountered in using spreadsheet formulas and the learning process towards achieving mastery in it. It consisted of four broad areas: discussion of participants' background (i.e. their work, daily tasks, and work tools); how they developed their expertise in spreadsheet use, and the facilitators and barriers to their general spreadsheet use experience. 

In addition, the interview guide consisted of a spreadsheet walk-through segment where participants provided example formulas, followed by an interactive segment, where they demonstrated how they would author one of those and compare the same workflow in Google Sheets and Microsoft Excel. Of particular relevance to our current research question was that the guide was designed to identify the specific spreadsheet practices and features used during collaborative work. In that sense, the Study 2 interview guide was supplementary to the Study 1 guide, which focused on abstract social contexts and influences underlying collaboration (See Appendix C for Study 2 interview guide).

\subsubsection{Procedure}
Participants were invited via email and asked to fill out the survey capturing their demographics. Interviews were between 1-1.5 hours long, conducted in-person, and recorded via audio recorders or Microsoft Teams. Each interview began with a discussion around the participant's background and how they developed their knowledge of spreadsheet formulas. Following the discussion, each participant was asked to walk us through their example formulas and demonstrate how they write or apply those to their work.

\subsection{Analysis}
\subsubsection{Positionality statement}
In data analysis, we adopted critical realism as an epistemological position \cite{braun2019, pilgrim2014}, and accordingly, we interpret the participants’ language as a reflection of their unique, personal experience of spreadsheet use, collaboration, and learning \cite{hall1997}. This is important as we were interested in the underlying beliefs, attitudes, and social norms that obstruct the willingness to engage in expertise sharing—but recognise that the participants themselves may not describe these with the same terms we do.

\subsubsection{Theory of planned behaviour (TPB)}

We used the theory of planned behaviour (TPB) to interpret the findings of Study 1 and Study 2. TPB explains social-related behaviours as intentional acts of decision-making, informed by three factors: attitude (expected outcomes of the behaviour), social norms (external expectations and motivations to conform to these norms), and perceived behavioural control (consisting of internal self-efficacy, and external constraints such as resource availability) \cite{ajzen1985, ajzen1991}. TPB is therefore well-suited for providing a comprehensive overview of volitional behaviours, such as knowledge sharing, and has been widely applied in knowledge management research for this purpose \cite{fauzi2024, nguyen2019}. For both of our studies, TPB helped to identify the individual, social, and technological factors influencing the intention to share spreadsheet-related expertise.

\subsubsection{Thematic analysis}
Both Study 1 and Study 2 interviews were treated as a single, complementary dataset, and analysed using reflexive thematic analysis. We applied both inductive and deductive approaches to qualitative coding by utilising the `template analysis' method established by King \cite{king1998}. `Templates' (i.e. pre-defined code structures) related to the core research goals were generated and applied a priori with the flexibility to expand with new codes as the analysis progressed. This was particularly useful as we needed to organise the coding template under TPB, while also ensuring that the additional codes were reflective of experiences and beliefs specific to the spreadsheet context.

We began by inductively coding the participants' descriptions of their interactions, beliefs, and experiences of using and sharing spreadsheets. The initial work of developing and refining the coding template was carried out by the first author. To ensure analytical rigour and reflexivity, the final template and a selected sample of quotes were shared with the second author, a senior researcher with prior experience in spreadsheet research. Their domain expertise provided alternative insights and served to triangulate the first author's interpretation \cite{guion2002}.

Once an agreement was reached between the authors, we followed a deductive approach and reorganised the codes according to the components of the TPB framework: attitudes around spreadsheet-related expertise sharing, subjective norms of spreadsheet-related expertise sharing, and perceived behavioural control over spreadsheet-related expertise sharing. Appendix E includes documentation of how the coding template was developed and refined over the course of analysis, and how the agreement was reached between the two authors. 

\section{Results}
In the following sections, we present our findings \comments{under themes by incorporating quotes from the Study 1 and Study 2 participants denoted S1P(no) and S2P(no) respectively.} We found \comments{that the} participants have competing attitudes towards sharing activities, and they deliberately determine when and how to share spreadsheet-related knowledge. Furthermore, these decisions are often influenced by perceived effort of communication, considerations of social timing and roles, conformity to perceived norms, and evaluation of one's own expertise and resources.

\subsection{Attitudes around spreadsheet-related expertise sharing}
\subsubsection{Spreadsheet recipients as a source of errors and disruption}
Participants had different, often conflicting attitudes towards spreadsheet sharing behaviour, depending on their role as providers or recipients. Among providers, spreadsheet sharing was commonly associated with frustrations, confusion from recipients, and an increased likelihood of errors. Recipients were often described as potential sources of changes disruptive to providers' goals. For instance, S2P3 noted: "\textit{Because all the stuff I do is touched by so many people [...] I don't want them to mess with it."} Participants who created their own spreadsheets anticipated that other users may struggle to understand and correctly implement changes according to their original intentions. An example is S1P4, who stated that "\textit{you've set up a bunch of different formulas and you know how they work and why they work. Someone else might not know.}"

As a result, participants intentionally limit input from others to minimise possible disruptions. Some restrict user rights, lock files, or even socially enforce their role as the sole editor or controller of the spreadsheet from the start. Continuing with their example, S1P4 provided an example of how they would protect a sheet that was crucial to their calculations by "\textit{lock[ing] the entire Excel file, because it's going to be used by a couple of people [...] if they accidentally hit something then things could break and it takes a long time to figure out why it's not working.}" 

\subsubsection{Knowledge sharing interactions provide users with significant support}
Despite the challenges faced by \comments{providers, recipients consider these individuals} useful sources of knowledge. Solutions, methods, and examples found in others' spreadsheets or in answers online could easily be reused for recipients' own purposes. S1P1 described such an instance where they took features from their colleagues' model: "\textit{So largely, I have used examples from other models. I've kind of patched some things together because some things work better than others.}" In some cases, recipients may inherit spreadsheet templates as examples from other colleagues, which they then develop according to their own preferences (S1P2, S1P26). 

The participants further acknowledged the usefulness of knowledge sharing interactions when they are the recipients. Providers’ advice and suggestions often prompted further discovery and, as was the case for S2P13, even increased motivation for learning: "\textit{There was a particular colleague who [...] just said something offhand about how much easier this would be to do in Excel, and from there, that kind of prompted me to then start asking questions of other people who I know use it, including this guy, to start using it more.}" S1P3 also reported that colleagues may re-share discoveries of useful features with others in their local community (e.g. via an internal Microsoft Teams channel). 

Some participants also engaged in proactive help-seeking from others, though this did not always lead to learning. `Local' providers (i.e. those with personal associations to the individual) were typically viewed as trusted sources of expertise, as S2P14 \comments{exemplified:} "\textit{Sometimes there will be more senior bankers sat around you that you go up to [to] ask, `I've got this problem. I've never encountered it before. In your years of experience, have you come across anything like this?'"} However, some participants developed more dependent relationships. In these instances, rather than developing their own skills, they may instead delegate more challenging tasks to others for support. For instance, S1P32 described the following: "\textit{I basically go, `Ooh, can't do this,' and I've fired it down for the spreadsheet person. They'll fix it for me, you know?}" S1P25 similarly explained that, when they encountered information which they knew was "\textit{above [their] head, [their] skill set, and [their] understanding}," they would "\textit{[do] all the bits [they] needed to do [...] and then ask somebody else to actually help [them] pull it through}."

\subsection{Subjective norms of spreadsheet-related \edit{expertise} sharing}
\subsubsection{\change{Familiarity with spreadsheets thought to be commonplace, downplaying the importance of \edit{expertise}}}
\label{subsubsec:spreadsheet_familiarity}
\comments{Participants frequently expressed} that `most' individuals would be familiar with using the spreadsheet, prompting unfounded assumptions about others' spreadsheet expertise. These views are largely driven by the intuitive nature of spreadsheet interactions, and a perception that spreadsheets \comments{are widely available and well established}. For instance, S1P32 stated: “\emph{Look, Excel, everybody knows how to use it […]. You just have to right click and add a new tab, and you're pretty much done.}” These beliefs often led participants to make assumptions about others' expertise, which often led to surprise when these assumptions are violated: "\emph{I'm always surprised when people don't know how to use formulas in Excel or automatically expand things or do conditional formatting}" (S1P26). Some participants, such as S1P24, expected greater proficiency from younger colleagues, and \comments{felt} surprised when this \comments{was} not the case: "\emph{[It’s] really surprising to me. A lot of younger people leaving school don't seem to be able to use Excel.}"

The belief that `most' people possess some form of spreadsheet expertise \comments{has} negative implications when reflected in organisational structures. Notably, it appears to close down opportunities for spreadsheet-related knowledge sharing. For example, S1P22 noted that they were unable to get answers to their queries about Microsoft Excel from their Information Technology (IT) department, jokingly stating that the department found their request to be out of their responsibility and “\textit{beneath them}.” An extreme example was given by S1P24, who, despite lacking technical knowledge, was forced to recover corrupted spreadsheet data from a virus because their organisation wanted to avoid outsourcing the issue to an external IT team.

Dismissive attitudes towards spreadsheet education, both in industry and in academic institutions, directly contribute to challenges in recruitment. S1P32 reflected: "\textit{When you find that person who knows Excel, they're your best friend for a lot of the day. But it's never something we look for when we're interviewing, bizarrely."} S1P21, who works in a treasury, similarly noted: "\emph{They [students] don't actually use Excel in [university] ever. […] it would be impossible to recruit students from university level who have the skills you need, it'd be 1 candidate out of 10,000}." As a result, spreadsheet training for new recruits (e.g. S1P21) is either the responsibility of existing managers or employees, or else recruits are simply asked to learn on the job. This, as S1P3 (a junior consultant) reflected, becomes "\textit{a little overwhelming at first, especially when you're thrust in a work environment and they throw a bunch of numbers at you and they're like `hey, do your best!'"}

In some cases, even expressing an interest in learning and using spreadsheets seems to trigger dismissive sentiments and negative judgments. The participants' statements indicate that it was socially acceptable to question those who dedicate time and effort to learning about and using spreadsheets. S1P29, a chemist, remarked: "\textit{I think we probably think of them [spreadsheets] as being a necessary evil. I do think there's some people that love them a bit too much.}" Further, even when positively recognising a colleague for solving a problem, S2P9 used the label "\textit{Excel nerd}." Individuals who expressed interest or excitement at the introduction of new spreadsheet functionalities could encounter surprise or skepticism from others. This was the case for S1P21: "\textit{When they introduced XLOOKUP [...] I was so excited and all my colleagues were like: `why? what's wrong with you?'"}

\subsubsection{\change{Personalised spreadsheet practices are valued, but there are pressures to conform to poorly defined `normal' practices}}
\label{subsubsec: personalised_pract}

Participants frequently described the spreadsheet's capacity to support personalised sensemaking practices \comments{as a great asset. Relatedly, participants acknowledged that the way spreadsheets are used varies across individuals.} "\textit{Every senior banker has a different style, as well, and they'll ask for deviation from the template,}" \comments{S2P14 commented when describing their work with senior colleagues. Participants also discussed their own personalisation needs} and appreciation for freedom and control that the spreadsheet offers. S1P22 likened the spreadsheet to "\textit{a big empty notebook [they] can use to [their] will.}" Similarly, S2P4 stated that, when using a spreadsheet, they can "\textit{organise everything in a way that makes the most sense to [them].}" 

\change{Despite these advantages, participants' accounts suggest \comments{that spreadsheet sharing in practice is often subject to certain expectations of standards and presentation—though the expectations may not be explicit.} "\textit{There's guidelines around what the data needs to look like, but not guidelines necessarily of how you need to do that and make that work,}" \comments{said S1P25 when describing the types of data that their organisation expected to receive from clients.} An awareness of these implicit expectations, and a desire to conform—or appear to conform—to these standards, mean that} \comments{the personalised elements of participants' spreadsheets often} pose a barrier to sharing. For example, S2P10 stated: "\textit{I wouldn't even really want to share the template for it, if the company asked me to provide it, because I feel this is very... it suits my needs.}" 

\subsubsection{\change{Sharing may not be appropriate depending on timing and individual roles}}
Even in a collaborative context, individual roles and responsibilities are typically varied. Some users have rigid expectations as to who needs to complete a particular task. As a result, collaboration is typically sequential, in that users carry out work only \textit{after} the previous user has completed theirs, rather than simultaneous, real-time discussion and collaboration. For instance,  S1P1 described: "\textit{Generally we avoid making changes [on others' spreadsheets] because if you have too many people doing things then you can lose track of how things are done."} This was also illustrated by S1P5, an IT consultant, describing the process of data cleaning with \comments{their} team: "\textit{They're [the colleagues] waiting for whatever data to come through, so I do need to be aware of that.}" 

The extent to which individuals feel they can rely on others is also influenced by their own roles and identity. For example, S1P25 felt compelled by their seniority to develop proficiency and independence: "\textit{As a head of department [...] there's things I can go and say to someone like the Data and Insights team, `I need this,' [...] but if I'm expecting my account managers and people working underneath me to be able to do these things that I feel I have to demonstrate it myself as well}."  

Some participants were restricted by data confidentiality and privacy policies in terms of which spreadsheets they could share or what aspects of their work they could discuss with colleagues. "\textit{Historically, [...] I would have been very cautious about who got a copy of a particular spreadsheet,}" S1P15 noted. As a consequence, exposure to others' spreadsheet practices or feature usage is limited, as S1P4 explained: "\textit{I don't know if I know of how other people in my industry still store spreadsheets, [...] we don't share any of this stuff.}"

\subsection{\edit{Perceived behavioural} control over \edit{spreadsheet-related expertise} sharing behaviours}
\subsubsection{Users tended to negatively evaluate their spreadsheet expertise}
Confidence in one's own knowledge is an important prerequisite to engage in knowledge sharing. However, some participants' statements showed they lack such confidence. For instance, S2P12 stated: "\textit{I can do basic formulas, but I'm not, you know, doing pivot tables and all that kind of thing.}" Some participants were overwhelmed by the extent of the spreadsheet software's features, assuming that they only accessed a fraction of its capabilities. S2P13 speculated: "\textit{There could be something out there that helps me with something, I'm just not aware of it.}" Similarly, S2P3, when asked to report on their own expertise, stated: "\textit{Just the fact that there are books of them means there's a million [formulas] out there that I know nothing about.}" Since the participants commonly believe that spreadsheet expertise should be widespread (see section \ref{subsubsec:spreadsheet_familiarity}), a lack of expertise is portrayed in a negative light. An example is S1P32, who described their own knowledge as "\textit{embarrassing}," stating that "\textit{you'd think after 20 something years my experience should be a little bit higher than it is.}"

Even the participants with confidence in their own knowledge—typically from finance or data science—noted with caution that their knowledge is restricted to their specific role. S2P10, a data scientist, stated: "\textit{At the level I think I need to be at, I'm very confident with what I do.}" S1P2, who works as an economics consultant, stated: "\textit{I feel I have very basic knowledge and specific knowledge of what I need to know, but aside from that, I would be clueless to do some basic things.}" This attitude stems from a lack of standardised definition for spreadsheet expertise, leading the participants to gauge expertise via various measurable attributes. These attributes range from comparisons with others' recall, speed, complexity of spreadsheet formulas to attributions based on individual characteristics such as age and profession. S2P3, for example, was impressed at their colleague's ability to memorise formulas: "\textit{[They] literally [go] to a cell, and [they type] the whole formula [...] And I’m like, ’how do you even know that stuff?'}" Similarly, S2P5 discussed how "\textit{younger guys in [their] office do a lot more funky stuff than [they] do.}" In some cases, individuals may classify the same spreadsheet features under different levels of difficulty, as S1P21 described: "\textit{To them [new recruits], `advanced' means pivot tables. And I would classify pivot tables and SUMIF, things like that, as basic.}"

This lack of clear standardisation led some participants to downplay the value of some beneficial spreadsheet practices. For example, when asked about their motivation for learning Visual Basic for Applications (VBA) and automating the functions within their spreadsheet, \comments{S1P21 stated that it is "\textit{mainly because by nature [they are] actually quite lazy."} As a result, S1P21 is motivated to find "\textit{any quicker, easier way of doing [things]}."} S2P13 similarly described their strategy of reusing spreadsheet templates or existing formula solutions online as "\textit{laziness}." Despite these points, both participants had self-reported high levels of experience in our surveys, suggesting that judgments of the value of spreadsheet expertise do not always correspond to the perceived depth of expertise.

\subsubsection{Effort is required to support effective communication with others}
When sharing, reviewing, and editing spreadsheets are unavoidable, participants often expend additional effort to modify, extract, and translate information in ways suited to the needs of others. In cases where a spreadsheet is complex, not all of the information in it is relevant for others. For instance, as S1P3 exemplified, "\textit{a lot of times, they'll [clients] be speaking to senior leadership and they don't have the time to look at your model, like never.}" As a result, some participants were averse to the idea of excessive sharing, preferring to provide summaries of the problems they encountered or the spreadsheet's overall output instead. For example, S2P13 shared an example of a department budget spreadsheet, which they "\textit{show[ed] to [their] director, so [they] don't include a lot of the other, smaller transactions.}"

Other participants noted how they would intentionally adapt their personal practices when anticipating that their spreadsheets would be shared with others. One experienced user, S1P21, explained how they would refrain from their usual practices during collaborations to accommodate others' expertise: "\textit{If I have a spreadsheet that I know is just for myself, I do tend to do it properly [...] whereas if I do the full belts and braces, [other users] get really stuck because there's too much for them to unpick.}" S1P45 similarly described how they would be more concerned with formatting, centering, and aligning their inputs when sharing with others compared to using spreadsheets on their own.  

In an extreme case, spreadsheet sharing \comments{was avoided} altogether because the participant \comments{did not want to spend time} explaining or cleaning data for others. S2P4, a teaching administrator, explained: "\textit{I don't really share [spreadsheets], it's not that I don't want to, it's [...] because then they [colleagues] would say: 'why did you do this?' And I just can't spend time explaining to them the reasons of why I do everything, because it would just be too time-consuming.}" 

\section{Discussion}

Overall, our findings reveal unique social barriers that spreadsheet users face in expertise sharing which require them to spend significant effort to address. While some of these barriers are driven by concerns of errors or miscommunication, we found that subjective standards of spreadsheet expertise and the timing of when sharing is considered appropriate also play a part. The design of spreadsheet software contribute\comments{s} to the participants' mixed self-evaluations of expertise and reinforces the assumption that spreadsheet skills are widespread, easy to acquire, and not explicitly valued in organisations. All these insights provide a comprehensive perspective of how sociotechnical factors influence spreadsheet expertise sharing.

In the following sections, we discuss how \comments{design aspects of spreadsheet software contribute to social norms influencing users' knowledge sharing intentions.} We contextualise our findings within previous literature on spreadsheet-based collaborations, and propose implications for design to promote greater knowledge sharing engagement. 

\subsection{\change{Spreadsheets as a platform for collaboration and knowledge sharing}}
\label{subsec: disc_personalised}
Our research challenges some of the pre-existing notions of how work is distributed between spreadsheet users. It is well-established that spreadsheets can support a diverse range of possibilities for accomplishing a task \cite{hermans2016}. \comments{Nardi and Miller's \cite{nardi1991} earlier work suggested that this quality provides opportunities for mutual learning, because it enables collaboration between technical experts (e.g. a competent spreadsheet user) and non-technical domain experts (e.g. a business owner) when solving domain-relevant tasks.} Our study, however, suggests that user collaborations are more sequential and distributed in nature. Users often alternate control and access over a spreadsheet, and there are unspoken expectations for when intervention from another user is considered acceptable. In some cases, only a single individual is given \comments{editing privileges} (e.g. S1P1), while other users are limited to commenting \comments{only}. Participants suggested that such turn-taking is necessary to avoid errors, miscommunication, and challenges around version control. \comments{These issues have been well-documented} in previous literature \cite{caulkins2007, panko2016}. However, it also limits users' exposure to learning new and alternative practices from others.

A further observation from our study suggests that individual differences in spreadsheet expertise can exacerbate problems with collaboration. In particular, we note that while users embrace the flexibility of the spreadsheet, they also view their own spreadsheet use strategies to be highly personalised and `non-standard,' though definitions of `standards' are often vaguely defined. \comments{This perception makes users reluctant to engage in sharing practices, as seen with participants such as S2P10 (see section \ref{subsubsec: personalised_pract}}), where their spreadsheet was too specific to their own needs to be sent to others. This desire to conform, or avoid highlighting one's `other-ness,' seems to be linked to concerns around self-presentation. These concerns were most evident when participants described their social roles and any accompanying expectations of responsibility (when in positions of seniority e.g. S1P25) or expertise (due to length of their career e.g. S1P32). Earlier works have demonstrated that the social relationship between providers and recipients, along with concerns of organisational surveillance, are important factors affecting user acceptance of expertise sharing systems \cite{pipek2012, giannisakis2022}. Our work further emphasises how subjectivity and differences in user practices can exacerbate these social concerns in knowledge sharing around software.

Furthermore, providers who are concerned about their `non-standard' spreadsheet practices will expend more effort to extract, filter, or condense their thought processes in a way which they believe would meet the needs of their recipients. This finding concurs with recent studies highlighting that authors are highly motivated to annotate and modify spreadsheets to facilitate collaboration \cite{chalhoub2022, srinivasaragavan2021}. Our work expands these studies by emphasising that user motivations are as much about conformity as they are about communication, and that this nuance limits user willingness to share. Smith et al. \cite{smith2017} had previously observed that experienced users may refrain from using advanced spreadsheet features in collaborative contexts, and we have found further evidence to corroborate this finding (e.g. S1P21). 

Overall, our study highlights how social concerns and motivations are deeply embedded in perceptions of spreadsheet interactions, to the extent that they interfere with collaborative behaviours. While normative beliefs about standard practices and concerns of self-presentation may make spreadsheet users reluctant to reveal their personal approaches, these tendencies \comments{may} be moderated if spreadsheet software provided greater ease for standardising and communicating such standards between users. Similarly, improving synchronous collaboration—for example, by offering support for tracking edits, user history, and version control—could promote further opportunities for knowledge sharing.

\subsection{\change{Feature-rich software and the challenge of knowledge self-efficacy}}
\label{subsec: knowledge_SE_spreadsheet} 
Within the knowledge management literature, knowledge self-efficacy (i.e. an individual's confidence in the value of their knowledge contribution) is an important pre-requisite for expertise sharing to take place \cite{nguyen2020, hung2015, kankanhalli2005}. As a theoretical construct, self-efficacy is informed by an individual's personal experiences of desirable or undesirable consequences, social comparison, vicarious observation of others, or verbal persuasion \cite{bandura1986}. Furthermore, it is captured in several behavioural theories commonly applied to knowledge sharing, such as social cognitive theory \cite{bandura1986} and TPB \cite{ajzen1985, ajzen1991}. Our study placed a particular focus on how interactions with the spreadsheet may produce experiences which heighten or diminish one's knowledge self-efficacy. In the context of spreadsheet software, knowledge self-efficacy specifically concerns how an individual self-reports their own spreadsheet expertise.

Our findings show that self-evaluations of spreadsheet expertise are often mixed. Most participants expressed confidence in their ability to apply spreadsheets to address work-related needs. Several participants remarked on the simplicity of spreadsheet use, and shared their belief that most would know how to use a spreadsheet. Yet many of the same individuals were also cautious to credit themselves as knowledgeable. We noted several instances where \comments{the} participants portrayed the spreadsheet as containing an insurmountable number of features, and framed their own expertise as a small proportion of the overall total.

These findings mirror a core problem previously identified in software learning research \cite{cockburn2014}. That is, users often fail to adapt advanced features and practices of feature-rich software even if they have long-term experience with using it \cite{nilsen1993, mcgrenere2000, mcgrenere2002, cockburn2014}. Part of the challenge, as discussed in section \ref{subsec: disc_personalised}, is that feature-rich software offer multiple ways to accomplish a task \cite{mcgrenere2000, nardi1991}. However, due to this multiplicity, the process of learning and deciding what novel features are relevant may become effortful according to previous studies \cite{kiani2019}.  This may explain why our participants tend to re-use existing solutions, rather than exploring novel features. Furthermore, analyses of corporate spreadsheets suggest that most users rely only on a small fraction of available formulas for the majority of tasks \cite{jansen2015, hermans2015}. This corresponds to our participants' self-deprecating descriptions of their own expertise.

The simultaneously conflicting remarks about the `simplicity' of spreadsheet use could be explained by the design of spreadsheet interfaces. Cockburn et al. \cite{cockburn2014} had previously suggested that intuitive user interfaces, designed with initial learnability in mind \cite{mcgrenere2000}, may unintentionally hinder users' discovery of advanced features and practices (e.g. hotkey shortcuts). While expert hotkey usage is somewhat more simplistic in nature compared to expert formula use in spreadsheets, some participants did indeed link the simplicity of spreadsheet use to its interface. Previous analyses also suggest that spreadsheet \comments{interfaces} incorporate several attributes conducive for initial learning and adoption, such as the grid, the taskbar, and \comments{their} immediate reactivity to user interactions \cite{maloney1995, hermans2016}. We do not necessarily argue that the current designs deter user discovery of more complex features (e.g. formulas, VBA). However, it is possible that alternative spreadsheet representations which provide greater visibility of underlying computational processes (see \cite{sarkar2018}) may encourage greater adoption of advanced features. 

In summary, we argue that the conflicted notions of spreadsheet expertise and self-efficacy observed may be attributed to two core factors. First, users are aware of the inherent complexity of the spreadsheet, and negatively evaluate their own expertise if they feel they are not sufficiently familiar with all its capabilities. Second, users may be disincentivised to explore and adopt more complex features (e.g. formulas), perhaps due to the intuitive and accessible nature of the spreadsheet's main interface. We suggest that these findings highlight the possible tension between balancing designs for initial learnability and the needs that arise in long-term learning \cite{sarkar2023}. Such considerations around software learnability directly affects the way users develop their knowledge self-efficacy, and thus are meaningful to consider for supporting knowledge sharing. However, further research is needed to establish whether feedback and framing interventions on self-efficacy will be effective for encouraging software-related knowledge sharing (e.g. \cite{nguyen2020, hung2015}).

Finally, while the role of knowledge self-efficacy in sharing is important in theory \cite{ajzen1985, bandura1986, nguyen2020, hung2015}, we acknowledge that the participants did not explicitly mention this concept in their discussions. This is partly related to the fact that interview data is dependent on interviewees' introspection abilities. Though an empirical relationship between spreadsheet knowledge self-efficacy and sharing has been identified in a recent study \cite{xia2024a}, \comments{we believe that further validation of the connection is necessary.}

\subsection{Implications for design}
\label{subsec:design_imp}
\edit{To address the challenges identified in section \ref{subsec: disc_personalised} and section \ref{subsec: knowledge_SE_spreadsheet}, we propose the following considerations for researchers and technology designers.}

\subsubsection{Supporting users in identifying relevant features}
\label{subsubsec:design_rel_feats}The large number of available features in spreadsheet software makes it challenging for participants to identify what is useful or relevant for their needs. We suggest that spreadsheet systems could address these issues by organising features based on their relevance to users' specific needs. This can reduce the amount of information users must consider when learning. Previous studies have explored adaptive interfaces that allow users to customise their preferred subset of features \cite{mcgrenere2007, mackay1991}, or tailor the taskbar according to users' preferences and use history automatically \cite{findlater2009, bunt2007}. Other studies have explored task-centric interfaces, such as `Workflows,' which organises available features based on specific tasks \cite{lafreniere2014a}. \comments{Some systems also} provide suggestions on how to carry out a task \cite{fraser2016, lafreniere2015}. These approaches may \comments{help} users to improve their proficiency, to notice differences between user preferences more easily, and therefore to discover previously unfamiliar features relevant to their work. 

An alternative approach to identifying features or software-related information is learning from other users in the same work domain. One source would be recommendations from colleagues or teammates \cite{giannisakis2022, bateman2013}, or members of a wider user community \cite{matejka2009, li2011, wang2018}. Previous studies have shown that this approach to learning is helpful for users to align their ways of use with others' \cite{giannisakis2022}. \comments{For spreadsheet users, standardisation and training are often severely lacking \cite{cragg1993, pemberton2000, lawson2009}, even for users within the same organisation \cite{smith2017}. Sharing knowledge between colleagues could help individuals develop mutually agreed standards for use.}

However, as we have previously discussed, designing systems to support expertise sharing requires consideration \comments{of two main aspects:} the quality of information extracted for recipients' benefits, \comments{and the} agency, control, and ownership of providers. As Giannisakis et al. \cite{giannisakis2022} suggest, passive sharing demands less effort but often causes a loss of control over shared information \cite{larsen-ledet2022, gausen2023}, while manual sharing and annotation demands more effort, but is desirable for users who are willing to support others. Future research that explores when and where an automated form of expertise sharing is most appropriate will be helpful. A careful consideration of how systems supporting such automation \comments{are} introduced into the specific culture of an organisation is necessary to ensure user acceptability and effectiveness.

\subsubsection{Improving clarity around user expertise}
An alternative approach for praxisware designers is to \comments{be more transparent regarding the inherent complexity of feature-rich software. This approach stands in contrast with the prevailing trend, which tends to favour simple designs and interfaces \cite{cockburn2014, carroll1987}. As Sarkar argues, users are both capable and willing to commit to long-term software learning when the benefits are made clear. Therefore, designers should acknowledge, rather than obscure, the challenges associated with such learning, and empower users throughout the learning process.} In line with Sarkar's argument \cite{sarkar2023}, our participants were quick to point out the necessity and value of the spreadsheet within their work despite its many established complexities, frustrations, and challenges \cite{panko2016, caulkins2007}. \comments{However, most participants struggled to articulate their technical understanding of the spreadsheet software or the standards of quality expected from others, which led to difficulties when self-evaluating expertise.} Exploring methods that offer clearer feedback and concrete performance indicators could help users better assess their expertise. This may reduce the perceived social costs of sharing software-related information, while encouraging further learning.

An example of such a feedback intervention is `Skill-o-meter' \cite{malacria2013}, a widget which provides visual feedback of the time taken to complete tasks for the user's chosen method (e.g. mouse-based interaction), and compares it to the time of using a recommended alternative (e.g. hotkeys). The use of Skill-o-meter was shown to increase the usage of hotkeys among users. \comments{A related} intervention called `Search Dashboard' applies a similar concept to web searches, allowing users to compare their own search history to `expert' searcher profiles and improve the efficiency of their own searching behaviours \cite{bateman2012}. Another related technique is labelling learning materials, such as tutorials and videos, according to varying levels of difficulty, which could also help users \comments{strengthen their understanding of} their own capabilities \cite{sabab2020}. 

Traditional metrics for expertise are often hierarchical or linear. This is thus challenging to apply in the spreadsheet context, where the most efficient method for a task may not necessarily be useful in another. For software used by diverse user groups, \comments{it may be more appropriate to acknowledge} different manifestations of expertise, rather than assign arbitrary value judgements and categorise users as either `low' or `high' expertise. Grossman and Fitzmaurice \cite{grossman2015} suggest using descriptive labels of expertise based on usage efficiency, familiarity with features, and relevance of knowledge to describe patterns in users' knowledge and preferences. For example, they describe `naïve' experts as users who lack knowledge of relevant alternative interaction methods but are highly familiar and efficient with a small subset of features, and `isolated' experts as users who may be efficient with a wide variety of interaction methods, but only for their specific domain purposes \cite{grossman2015}. This approach may help users to acknowledge the validity of their preferred practices, while providing clarity as to how other methods may help meet their needs. \comments{Designers could also delineate different spreadsheet use strategies based on the values important to users, such as rating or providing measurements of formulas based on readability \cite{hermans2012a} and weakness to errors. This provides users with practical criteria with which they could self-evaluate expertise rather than arbitrary and irrelevant impressions.} 

\section{Limitations}
\comments{We recognise that data analysis and interpretation under an interpretivist stance may possibly be subject to the researcher's personal experiences and values \cite{blandford2016, madill2000}. To mitigate this, we took measures where possible. We engaged in the principles of reflexivity and introduced alternative perspectives \cite{braun2019} from a second interview dataset in data analysis. While the first author was ultimately responsible for the final interpretation and presentation of these findings, using data collected by different researchers (non-authors) and incorporating discussions with co-authors provided a more rounded perspective in analysis.}

\comments{The other limitation} was that we did not collect demographic data concerning participants' cultures, a potential factor which can influence expertise sharing practices \cite{ardichvili2006}. \comments{However, since most participants were recruited in the United Kingdom, it may be assumed that we do not have} sufficient sample size to identify potential cultural trends. This issue could be explored further in future research. While the interviews were conducted \comments{by us, which allowed us} to validate participant engagement, we did not administer attention checks during the survey, \comments{which limited the validity of responses.}

\section{Conclusion}
\comments{We provide two contributions in this paper. The first contribution is our investigation into the social barriers around spreadsheet expertise sharing. Through a qualitative analysis of interviews with 31 spreadsheet users, we identify how users struggle to adapt personalised use strategies to subjective standards of quality and to navigate timings for sharing. We also find that concerns over potential collaboration errors, conflicted self-evaluations of spreadsheet expertise, and wider social norms devaluing spreadsheet-related knowledge, further disincentivise expertise sharing practices. We highlight how these challenges are inextricably linked to the flexibility and complexity of feature-rich software. Our second contribution is a set of design recommendations based on our findings which may help address the challenges identified. We suggest that different design choices can support users in developing realistic expectations for expertise by: being more transparent about the complexity of using a feature-rich software, using taskbars which support personalisation and prioritise feature relevance, and emphasising descriptive labels of expertise over hierarchical labels. The findings of this paper suggest that considerations for alternative design choices may lead to different user dynamics and engagement in collaborative behaviours in the long term. By highlighting the sociotechnical nature of expertise sharing practices, our paper provides insights and reflections which may be useful not only for spreadsheet users, but could also inform future software design considerations.} 

%% The next two lines define the bibliography style to be used, and
%% the bibliography file.
% \bibliographystyle{ACM-Reference-Format}
% \bibliography{CSCW03_2025}
%%% -*-BibTeX-*-
%%% Do NOT edit. File created by BibTeX with style
%%% ACM-Reference-Format-Journals [18-Jan-2012].

\end{document}